\documentclass{ifacconf}
\usepackage{color}
\usepackage{amsmath}
\usepackage{mathtools}
\usepackage{amsfonts}
\usepackage{graphicx}      % include this line if your document contains figures
\usepackage{natbib}        % required for bibliography
\newcommand{\iu}{{i\mkern1mu}}
\graphicspath{{./figs/}}

\begin{document}
\begin{frontmatter}

\title{Hierarchical Fault-Tolerant Coverage Control for an Autonomous Aerial Agent} 

\thanks[footnoteinfo]{This work was undertaken as part of the GLIMPSE project EXCELLENCE/0421/0586 which is co-financed by the European Regional Development Fund and the Republic of Cyprus through the Research and Innovation Foundation's RESTART 2016-2020 Programme for Research, Technological Development and Innovation and supported by the European Union's Horizon 2020 research and innovation programme under grant agreement No 739551 (KIOS CoE), and from the Government of the Republic of Cyprus through the Cyprus Deputy Ministry of Research, Innovation and Digital Policy.}

\author{Savvas~Papaioannou,} 
\author{Christian~Vitale,} 
\author{Panayiotis~Kolios,}
\author{Christos~G.~Panayiotou, and}
\author{Marios~M.~Polycarpou}

\address{KIOS Research and Innovation Centre of Excellence (KIOS CoE) and Department of Electrical and Computer Engineering, University of Cyprus, Nicosia, 1678, Cyprus (e-mail: \{papaioannou.savvas, vitale.christian, pkolios, christosp, mpolycar\}@ucy.ac.cy).}

\begin{abstract}                % Abstract of 50--100 words
Fault-tolerant coverage control involves determining a trajectory that enables an autonomous agent to cover specific points of interest, even in the presence of actuation and/or sensing faults. In this work, the agent encounters control inputs that are erroneous; specifically, its nominal controls inputs are perturbed by stochastic disturbances, potentially disrupting its intended operation. Existing techniques have focused on deterministically bounded disturbances or relied on the assumption of Gaussian disturbances, whereas non-Gaussian disturbances have been primarily been tackled via scenario-based stochastic control methods. However, the assumption of Gaussian disturbances is generally limited to linear systems, and scenario-based methods can become computationally prohibitive. To address these limitations, we propose a hierarchical coverage controller that integrates mixed-trigonometric-polynomial moment propagation to propagate non-Gaussian disturbances through the agent's nonlinear dynamics. Specifically, the first stage generates an ideal reference plan by optimising the agent's mobility and camera control inputs. The second-stage fault-tolerant controller then aims to follow this reference plan, even in the presence of erroneous control inputs caused by non-Gaussian disturbances. This is achieved by imposing a set of deterministic constraints on the moments of the system's uncertain states.
\end{abstract}


\end{frontmatter}
%==============================================================================

\section{Introduction} \label{sec:intro}

Coverage planning in autonomous systems (\cite{Tan2021,Papaioannou2022CDC,Papaioannou2023CDC}) refers to the problem of designing a trajectory or a series of actions for an autonomous agent to ensure that it covers a specified area or a set of points of interest, while at the same time optimising a particular mission objective. 

Recently, unmanned aerial vehicle (UAV) -based systems have been increasingly applied to various application domains as shown in \cite{ZachariaMED,PapaioannouMED,Soleymani2023,PapaioannouIROS2021} including  coverage control, as highlighted in \cite{Cabreira2019s,PapaioannouTAES,PapaioannouICUAS2022}. However, the majority of these approaches, primarily focus on covering 2D planar areas. These methods often employ simple geometric patterns, such as back-and-forth or zig-zag motions, which are not easily adaptable to 3D environments, and predominantly utilize UAVs with fixed, non-controllable sensors. 

In addition, in specific application domains, such as emergency response and critical infrastructure inspection, fault-tolerant coverage control (FTCC) is imperative. The primary goal of fault-tolerant control (FTC) is to sustain system performance close to the desired level despite various types of faults, including actuation and sensing faults. Over the last years, the field of FTC has gained considerable attention (\cite{zhang2008bibliographical}). Numerous FTC methods, including both passive and active approaches, have been proposed in the literature (\cite{jiang2012fault}) for linear and nonlinear systems to address a variety of faults. However, several challenges remain unresolved, particularly in dealing with faults arising from random disturbances, such as erroneous control inputs caused by stochastic and potentially non-Gaussian disturbances.

Existing methodologies usually employ deterministically bounded disturbances (\cite{mayne2005robust}) or are based on linear-Gaussian assumptions regarding the disturbances affecting the system (\cite{vitale2022autonomous,dai2019chance}). While fault-tolerant and robust methods accommodating nonlinear system dynamics (\cite{lew2020chance}) have been developed, they often presuppose Gaussian disturbances \cite{Papaioannou2023unscented}. In contrast, to handle non-Gaussian disturbances, scenario-based stochastic control methods have been introduced (\cite{summers2018distributionally,Calafiore2013}), however, these are generally constrained to linear systems. 

To address these limitations, we propose in this work a fault-tolerant hierarchical coverage controller. It accounts for stochastic non-Gaussian disturbances affecting the nominal control inputs of an autonomous UAV agent. Consequently, these disturbances  propagate through the agent's nonlinear dynamical model, affecting its desired operation. The proposed approach allows for the accommodation of these disturbances, thereby facilitating the generation of fault-tolerant coverage plans. 

\section{Preliminaries} \label{sec:model}

\subsection{Agent Dynamical Model} \label{ssec:agent_dynamics}
In this study, we focus on an autonomous UAV agent whose state at time-step $t$ is represented by $\boldsymbol{x}_t \in \mathcal{X} \subset \mathbb{R}^4$. This agent operates within a bounded 3D surveillance area, $\mathcal{A} \subset \mathbb{R}^3$, and follows the stochastic discrete-time dynamical model $\boldsymbol{x}_t = f(\boldsymbol{x}_{t-1},\boldsymbol{u}_{t-1},\boldsymbol{\omega}_{t-1})$, outlined below:
\begin{subequations} \label{eq:nonlin_dynamics}
\begin{align}
	x_t &= x_{t-1} + \Delta_t(u^\nu_{t-1}+\omega^\nu_{t-1})\cos(\psi_{t-1}),\\
	y_t &= y_{t-1} + \Delta_t(u^\nu_{t-1}+\omega^\nu_{t-1})\sin(\psi_{t-1}),\\
	z_t &= z_{t-1} + \Delta_t(u^z_{k-1}+\omega^z_{k-1}),\\
	\psi_{t} &= \psi_{t-1} + \Delta_t(u^\psi_{t-1}+\omega^\psi_{t-1}).
\end{align}
\end{subequations}
\noindent Here, $\boldsymbol{x}_t = [x_t, y_t, z_t, \psi_{t}]^\top$ represents the agent's position $(x_t, y_t, z_t)$ and orientation $(\psi_t)$, i.e., yaw, in 3D Cartesian coordinates. The control input vector $\boldsymbol{u}_t = [u^\nu_t, u^z_t, u^\psi_t]^\top \in \mathcal{U}$ comprises the horizontal and vertical velocities $(u^\nu_t, u^z_t)$, and the yaw rate $(u^\psi_t)$, and $\Delta_t$ denotes the sampling interval. The term $\boldsymbol{\omega}_t = [\omega^\nu_t, \omega^z_t, \omega^\psi_t]^\top$ represents random disturbances acting on the agent's nominal control inputs, thereby affecting its normal operation. 

\subsubsection{Assumption 1:} The disturbance term $\boldsymbol{\omega}_t$ is a probabilistic process noise vector consisting of independent components, i.e., random variables, which: a) can follow different probability density functions (potentially non-Gaussian), and b) have finite moments of order $\alpha \in \mathbb{N}$.

\subsection{Agent Sensing Model} \label{ssec:agent_sensing}

The UAV agent is outfitted with a gimballed camera exhibiting a limited field-of-view (FOV), and rotation capabilities. The camera FOV is modelled as a regular right pyramid with four triangular sides and a rectangular base. The optical center of the camera, is located at the pyramid's apex, directly above the FOV base's centroid. The dimensions of the FOV's rectangular base, namely its length and width are denoted as $\ell_f$, and  $w_f$ respectively, and the camera's observation range is denoted as $h_f$ (i.e., pyramid's height). The camera's initial pose (i.e., the vertices of the FOV) are given by the 3-by-5 matrix $C_0$, as follows:
\begin{equation} \label{eq:C0}
    C_0 =
    \begin{bmatrix}
       h_f & h_f & h_f  & h_f &0 \\
       w_f/2 & w_f/2 & -w_f/2 & -w_f/2 &0 \\
        \ell_f/2  & -\ell_f/2  &  -\ell_f/2  &  \ell_f/2  &0 \\
    \end{bmatrix},
\end{equation}
\noindent where the camera's optical axis i.e.,  the line connecting the base's centroid and the pyramid's apex, of length $h_f$, aligns parallel to the $zy$-plane's normal. At time-step $t$ the camera's FOV can be rotated through two sequential rotations i.e., firstly, a rotation by an angle $\phi^y$ around the $y$-axis, and secondly, a rotation $\phi^z$ around the $z$-axis as: $C^i_t(\phi^y,\phi^z) = R_z(\phi^z) R_y(\phi^y) C^i_0, \forall i \in \{1,..,5\}$,
%\begin{equation}\label{eq:FOV_rotation}
%    C^i_t(\phi^y,\phi^z) = R_z(\phi^z) R_y(\phi^y) C^i_0, \forall i \in \{1,..,5\},
%\end{equation}
\noindent where $C^i_0$ denotes the $i_\text{th}$ column of the matrix $C_0$, $C^i_t(\phi^y,\phi^z)$ refers to the FOV vertex after rotation, as parameterized by the angles $\phi^y$ and $\phi^z$. The matrices $R_y(a)$ and $R_z(a)$ denote the fundamental 3-by-3 rotation matrices, which are used to rotate a vector by an angle $a$ around the $y$-axis and $z$-axis, respectively. Finally, the pair of rotation angles $\{\phi^y,\phi^z\} \in \mathcal{M}$ is taken from the finite set $\mathcal{M}$ of all  admissible camera rotations (i.e., pairwise combination of angles). In other words, the camera FOV can acquire at each time-step $t$, $1$ out of $|\mathcal{M}|$ possible configurations.

\subsection{Points of Interest}
In this study, we consider the surface area $\partial O$ of an object of interest $O$, to have been triangulated  into a 3D mesh $\mathcal{K} = \{K_1,..,K_{|\mathcal{K}}|\}$ composed of a finite number $|\mathcal{K}|$ of triangular facets $K \in \mathbb{R}^{3 \times 3}$, where $|\cdot|$ signifies the set cardinality. The points of interest, i.e., the points that need to be covered by the agent's camera, are defined as the centroid points of a specified subset of facets $\tilde{\mathcal{K}} \subseteq \mathcal{K}$ on the object's surface area. The centroid point of facet $K \in \tilde{\mathcal{K}}$ will be denoted hereafter as $\boldsymbol{c}_K \in \mathbb{R}^3$, and the set of all centroid points to be covered as $\tilde{\mathcal{K}}_c$.

\section{Robust Hierarchical Coverage Control} \label{sec:approach}

The problem tackled in this work can be stated as follows:
\textit{Given a rolling finite planning horizon of length $T$ time-steps, find the UAV's joint mobility i.e., $\boldsymbol{u}_{t+\tau|t}$, and camera i.e., $\{\phi^y, \phi^z\}_{t+\tau|t}$ control inputs for $\forall \tau \in \{0,..,T-1\}$ which maximise the coverage of the points of interest $\boldsymbol{c}_K \in \tilde{\mathcal{K}}_c$ on the object's surface area.}

To address this problem, it's important to recognize that the agent's state, $\boldsymbol{x}_t$, is a multivariate random variable due to random disturbances affecting the system. This uncertainty extends to the state of the camera's Field of View (FOV), making the coverage task inherently non-deterministic. To tackle this challenge robustly, we aim to ensure that the points of interest are covered with a certain probability $1-\epsilon$. 

%This goal necessitates the development of a controller capable of accommodating potentially non-Gaussian or unbounded process noise through the agent's nonlinear dynamics. This capability is crucial for calculating the probability distribution of the agent's state over the planning horizon. As a result, it facilitates the generation of fault-tolerant coverage plans, which guarantee coverage at specific confidence levels even under erroneous control inputs

In this section, we show how we have addressed the above challenges by designing a hierarchical fault-tolerant coverage controller. In the first stage (i.e., Sec. \ref{ssec:stage1}), the problem is reformulated as a disturbance-free rolling horizon model predictive control problem (MPC), generating finite-length look-ahead coverage plans. In the second stage (i.e., Sec. \ref{ssec:stage2}), a fault-tollerant controller is developed, employing exact uncertainty propagation techniques to propagate the disturbance through the agent's dynamics. This allows for the computation of constraints on the moments of the agent's uncertain states, ensuring fault-tolerant coverage control at specified probability levels.

\subsection{Stage-1: Coverage Reference Plan} \label{ssec:stage1}
The first-stage controller, as formulated in \eqref{eq:P2}, is designed as a rolling horizon MPC. In this setup, the agent computes finite-length look-ahead coverage trajectories $\boldsymbol{x}^1_{t+\tau+1|t}$ for $\tau \in \{0,..,T-1\}$. 
%Here, the notation $x_{\tau|t}$ represents the predicted state of the agent at future time-step $\tau \geq t$, as calculated at the current time-step $t$. This coverage planning process is executed iteratively across multiple time-steps $t$, where the first set of predicted control inputs $\boldsymbol{u}^1_{t|t} \in \mathcal{U}^1$ in the sequence is executed at the subsequent time-step. Following this, the agent transitions to its new state, and the optimization cycle described above is repeated for the next time-step, now over a shifted planning horizon of $T^1$ time-steps.
\begin{algorithm}[h]
\begin{subequations} \label{eq:P2}
\begin{align} 
%&\hspace*{-3mm}\texttt{Stage-1 MPC Controller} & \notag\\
& \hspace*{-2.5mm} ~~\underset{\{\boldsymbol{u^1}, \boldsymbol{b^{GC}} \}}{\min} ~\mathcal{J}^{1} = \mathcal{J}^{1}_\text{GC} +  w\mathcal{J}^{1}_\text{CE} &  \hspace*{-5mm} \label{eq:P2_0} \\
&\hspace*{-3mm}\textbf{subject to: $\tau \in \{0,..,T-1\}$} ~  &\nonumber\\
&\hspace*{-3mm} \boldsymbol{x}^1_{t+\tau+1|t} = \Phi \boldsymbol{x}^1_{t+\tau|t} + \Gamma \boldsymbol{u}^1_{t+\tau|t}, & \hspace*{-5mm} \forall \tau \label{eq:P2_1}\\
&\hspace*{-3mm} \boldsymbol{x}^1_{t|t} = \boldsymbol{x}^1_{t|t-1},&\label{eq:P2_2}\\
&\hspace*{-3mm} b^{\text{G}}_{\kappa,t+\tau+1|t} = 1 \iff  \text{pos}(\boldsymbol{x}^1_{t+\tau+1|t}) \in  \bigtriangleup(S_\kappa), & \hspace*{-7mm} \forall \kappa, \tau \label{eq:P2_3}\\
&\hspace*{-3mm} \sum_\tau b^{\text{G}}_{\kappa,t+\tau+1|t} \leq 1 , & \hspace*{-7mm} \forall \kappa \label{eq:P2_3b}\\
&\hspace*{-3mm} \hat{b}^{\text{G}}_{\kappa,t+\tau+1|t} \leq  b^{\text{G}}_{\kappa,t+\tau+1|t} + V(\kappa), & \hspace*{-7mm} \forall \kappa, \tau \label{eq:P2_4}\\
&\hspace*{-3mm} C_{m} =  R_z(\phi^z)R_y(\phi^y)C_0,  & \hspace*{-15mm} \forall \{\phi^z,\phi^y\}\in \mathcal{M}\label{eq:P2_5}\\
&\hspace*{-3mm} \hat{C}_{m,t+\tau+1|t} = C_{m}+ \text{pos}(\boldsymbol{x}^1_{t+\tau+1|t}), & \hspace*{-5mm} \forall m, \tau \label{eq:P2_6}\\
&\hspace*{-3mm} b^{\text{FOV}}_{\kappa,m,t+\tau+1|t} = 1 \iff  \boldsymbol{c}_K \in  \bigtriangleup(\hat{C}_{m,t+\tau+1|t}), & \hspace*{-7mm} \forall \kappa, m, \tau \label{eq:P2_7}\\
&\hspace*{-3mm} \sum_\kappa \sum_{m} b^{\text{FOV}}_{\kappa,m,t+\tau+1|t} \leq 1, & \hspace*{-7mm} \forall \tau \label{eq:P2_8}\\
&\hspace*{-3mm} \hat{b}^{\text{FOV}}_{\kappa,m,t+\tau+1|t} \leq b^{\text{FOV}}_{\kappa,m,t+\tau+1|t} + V(\kappa), & \hspace*{-7mm} \forall \kappa, m, \tau \label{eq:P2_8b}\\
&\hspace*{-3mm} b^{GC}_{\kappa,m,t+\tau+1|t} = \hat{b}^{\text{G}}_{\kappa,t+\tau+1|t} \wedge \hat{b}^{\text{FOV}}_{\kappa,m,t+\tau+1|t}, & \hspace*{-7mm} \forall \kappa, m, \tau \label{eq:P2_9}
%&\hspace*{-3mm}  b^{GC}_{\kappa,\hat{m},t+\tau+1|t}, b^{\text{G}}_{\kappa,t+\tau+1|t}, \hat{b}^{\text{G}}_{\kappa,t+\tau+1|t} \in \{0,1\} & \hspace*{-7mm} \forall \kappa, \tau \\ 
%&\hspace*{-3mm}  b^{\text{FOV}}_{\kappa,\hat{m},t+\tau+1|t}, \hat{b}^{\text{FOV}}_{\kappa,\hat{m},t+\tau+1|t}, V(\kappa) \in \{0,1\} & \hspace*{-7mm} \forall \kappa, \tau, \hat{m}\\ 
%&\hspace*{-3mm}  \hat{m} \in \{1,..,|\hat{\mathcal{M}}|\}, \kappa \in \{1,..,|\mathcal{\tilde{K}}_c|\} &\\
%&\hspace*{-3mm} \boldsymbol{x}^1_{t|t}, \boldsymbol{x}^1_{t+\tau+1|t} \in \mathcal{X}^1,~ \boldsymbol{u}_{t+\tau|t}^1 \in \mathcal{U}^1 & \hspace*{-7mm} \forall \tau
\end{align}
\end{subequations}
\end{algorithm}
In \eqref{eq:P2}, we generate an ideal coverage reference plan based on simplified linear dynamics of the agent in a disturbance free setting, as shown in \eqref{eq:P2_1}, where the matrices $\Phi$ and $\Gamma$ are given by:
\begin{equation}
\Phi = 
\begin{bmatrix}
    \boldsymbol{1}_{3\times3} & \Delta_t^1\boldsymbol{1}_{3\times3}\\
    \boldsymbol{0}_{3\times3} &  \boldsymbol{1}_{3\times3}
   \end{bmatrix}, ~
\Gamma = 
\begin{bmatrix}
    \boldsymbol{0}_{3\times3} \\
    \boldsymbol{1}_{3\times3}
   \end{bmatrix}.
\end{equation}

\noindent The matrices  $\boldsymbol{1}_{3\times3}$ and $\boldsymbol{0}_{3\times3}$ denote the 3-by-3 identity and zero matrices respectively, and $\Delta_t^1$ denotes the sampling rate. The agent's state $\boldsymbol{x^1_{t}}$ is composed of position and velocity components in 3D Cartesian coordinates, and the control input vector $\boldsymbol{u^1_{t}}$ represents the velocity components in 3D coordinates. 
\subsubsection{Assumption 2:} For each point of interest $\boldsymbol{c}_K \in \tilde{\mathcal{K}}_c$, there is an associated region of interest $S_\kappa \subset \mathcal{A}$ (where $\kappa$ is the index used to identify $\boldsymbol{c}_K$). When the agent resides within this region (i.e., $\text{pos}(\boldsymbol{x}_{t}) \in S_\kappa$), there exist at least one configuration of the camera FOV $m \in \{1,..,|\mathcal{M}|\}$ that can ensure the coverage of $\boldsymbol{c}_K$. $S_\kappa$ is identified as a sphere with radius $s_r$ centered at $\boldsymbol{\bar{s}}_\kappa = \ell\boldsymbol{a}^S_\kappa+\boldsymbol{c}_K$, where $\ell$ is a appropriately selected distance, and $\boldsymbol{a}^S_\kappa$ is the outward unit normal vector to the facet $K \in  \tilde{\mathcal{K}}$ which has as its centroid point, the point of interest $\boldsymbol{c}_K$. 

\textbf{UAV guidance through the regions of interest:} We utilize the binary variable $b^{\text{G}}_{\kappa,t+\tau+1|t}$ in \eqref{eq:P2_3} to identify whether the agent's position $\text{pos}(\boldsymbol{x}^1_{t+\tau+1|t})$ at the time-step $t+\tau+1|t$ inside the planning horizon (hereafter abbreviated as $\tau$) resides inside the convex-hull of the $\kappa_\text{th}$ sphere $S_\kappa$, associated with the point of interest $\boldsymbol{c}_K$. To implement this functionality with linear inequalities, we first linearise $S_\kappa$ by approximating it with the inscribed dodecahedron as in \cite{PapaioannouTMC2023}. Subsequently, we can identify the presence of the agent inside the dodecahedron as shown below:
\begin{subequations}\label{eq:v0}
\begin{align} 
&  \text{dot}(\boldsymbol{a}_{i,\kappa},\text{pos}(\boldsymbol{x}^1_{\tau})) + \bar{b}^{\text{G}}_{i,\kappa,\tau}(M-\beta_{i,\kappa}) \le M, \label{eq:v1}\\
& 12 b^{\text{G}}_{\kappa,\tau} - \sum_{i=1}^{12}\bar{b}^{\text{G}}_{i,\kappa,\tau} \le 0 \label{eq:v2}
\end{align}
\end{subequations}
\noindent where $\text{dot}(\boldsymbol{a}_{i,\kappa},\boldsymbol{x})=\beta_{i,\kappa}$ describes the equation of the plane which contains the $i_\text{th}$ facet of the $\kappa_\text{th}$ dodecahedron associated with the point of interest $\boldsymbol{c}_K$, $\boldsymbol{x} \in \mathbb{R}^3$, $\text{dot}(a,b)$ is the dot product between the vectors $a$ and $b$, and $M$ is a sufficiently large positive constant that makes sure that \eqref{eq:v1} remains valid for any value of the binary decision variable $\bar{b}^{\text{G}}_{i,\kappa,\tau} \in \{0,1\}, \forall i \in \{1,..,12\}, \kappa, \tau$. 
%Therefore, when the agent positional vector $\text{pos}(\boldsymbol{x^1_{\tau}})$ resides inside the convex-hull, of the $\kappa_\text{th}$ dodecahedron at time-step $\tau$ then the binary variable $\bar{b}^{\text{G}}_{i,\kappa,\tau}=1, \forall i$. Subsequently, the binary variable $b^{\text{G}}_{\kappa,\tau}$ can be utilized as shown in \eqref{eq:v2} to drive the agent inside the region of interest $S_\kappa$ at time-step $\tau$ (i.e., $b^{\text{G}}_{\kappa,\tau}=1$).
The constraint expressed in \eqref{eq:P2_3b} ensures that the agent generates plans that visit each region of interest exactly once within the planning horizon, and subsequently, the constraint in \eqref{eq:P2_4} makes sure that the agent avoids the duplication of work. If the agent has visited the region of interest $S_\kappa$ at some point in time, as indicated by $V(\kappa) \in \{0,1\}$ (i.e., a database that records the visited regions of interest), then according to \eqref{eq:P2_4}, there is no incentive to activate the variable $b^{\text{G}}_{\kappa,\tau}$ for the same region at a future time-step. 
%Consequently, the agent generates coverage plans focusing on unvisited regions, thereby avoiding duplication of effort

\textbf{FOV control:} In this stage, to determine the camera rotations that allows the coverage of the points of interest $\boldsymbol{c}_\kappa, \forall \kappa  \in \{1,..,|\tilde{\mathcal{K}}_c|\}$, we utilize constraint in \eqref{eq:P2_5} in order to pre-compute the camera FOV states $C_{m}$ for all possible FOV configurations $m \in \{1,..,|\mathcal{M}|\}$.  Subsequently, \eqref{eq:P2_6} translates at time-step $\tau$ the camera FOV state $\hat{C}_{m,t+\tau+1|t}$ at the predicted agent position i.e., $\text{pos}(\boldsymbol{x}^1_{t+\tau+1|t})$. Then, the binary variable $b^{\text{FOV}}_{\kappa,m,t+\tau+1|t}$ in \eqref{eq:P2_7} is activated when the point of interest $\boldsymbol{c}_K$ (indexed by $\kappa$) resides within the convex-hull of the $m_\text{th}$ camera FOV state at time-step $\tau$. This functionality is implemented similarly to \eqref{eq:v0} accounting for the 5 facets of the camera FOV. Constraint, \eqref{eq:P2_8} makes sure that only one FOV configuration is active at any point in time, constraint \eqref{eq:P2_8b} discourages the coverage of already covered points, and finally the logical conjunction in \eqref{eq:P2_9} integrates the guidance and coverage decision variables. In order, to generate the coverage plan over the rolling planning horizon we simultaneously optimize guidance and coverage as $\mathcal{J}^{1}_{GC} = - \sum_{\kappa=1}^{|\tilde{\mathcal{K}}_c|} \sum_{m=1}^{|\mathcal{M}|} \sum_{\tau=0}^{T-1} b^{GC}_{\kappa,m,t+\tau+1|t}$, as shown in \eqref{eq:P2_0}. The control effort denoted as $\mathcal{J}^{1}_{CE}$ is defined as $\sum_{\tau=0}^{T-1} ||\boldsymbol{u}^1_{t+\tau|t}||^2_2$, and $w$ is a tuning parameter.

In essence, the stage-1 controller generates the agent's trajectory $\{\boldsymbol{x}^1_{t+1|t},..,\boldsymbol{x}^1_{t+T|t}\}$, and a set of camera views $\{\hat{C}_{m_1,t+1|t},...,\hat{C}_{m_n,t+T|t}\}, m_i \in \{1,..,|\mathcal{M}|\}$ that maximize the coverage of the points of interest inside the planning horizon. We should note here that the agent's trajectory determines the order in which the various points of interest will be covered (i.e., a list which contains the sequence of coverage events obtained via $b^{GC}_{\kappa,t+\tau+1|t}$). The first element in this list (i.e., the point of interest which has been planned to be covered next, and the associated region of interest), is passed as input to the stage-2 controller.
\subsection{Stage-2: Fault-Tolerant non-Gaussian Control} \label{ssec:stage2}

At each time-step, the stage-2 controller receives from stage-1 the region of interest $S_\kappa$ for some $\kappa \in \{1,..,|\tilde{\mathcal{K}}_c|\}$, which is planned to be visited next. Subsequently, the objective of the stage-2 controller is to generate a fault-tolerant plan guiding the agent robustly within $S_\kappa$. Such objective is attained by optimizing the agent's control inputs $\boldsymbol{u}_{t+\tau|t}$ according to \eqref{eq:nonlin_dynamics} over a suitably chosen planning horizon i.e., $\tau \in \{0,..,T^\prime-1\}$ of length $T^\prime$ time-steps. To achieve this goal, we employ exact uncertainty propagation techniques of mixed-trigonometric-polynomial functions (\cite{han2022non}). 
%This approach enables us to design and impose deterministic constraints on the moments of the system's uncertain states, thereby effectively accommodating the random disturbances acting on the system's nominal control inputs with a specified level of confidence.

\subsubsection{Lemma 1:}
Consider a random variable $x$ with the characteristic function denoted by $\Phi_x(t)$. For a given triplet $(\alpha_1,\alpha_2,\alpha_3) \in \mathbb{N}^3$, where $\alpha = \sum_{i=1}^3 \alpha_i$, and a scaling factor $\delta$, the mixed-trigonometric-polynomial moment of order $\alpha$ of the form $\mu_{\delta,x^{\alpha_1} c^{\alpha_2}_x s^{\alpha_3}_x}$, is defined as $\mathbb{E}[(\delta x)^{\alpha_1} \cos^{\alpha_2}(\delta x) \sin^{\alpha_3}(\delta x)]$, and can be computed as:  
 \begin{equation}\label{eq:lemma_1}
\begin{aligned}
\mu_{\delta,x^{\alpha_1} c^{\alpha_2}_x s^{\alpha_3}_x}= & \frac{1}{\iu^{(\alpha_1+\alpha_3)}2^{(\alpha_2+\alpha_3)}} \sum_{g=0}^{\alpha_2}\sum_{h=0}^{\alpha_3}\binom{\alpha_2}{g}\binom{\alpha_3}{h}\\
 & (-1)^{(\alpha_3-h)}\frac{\partial^{\alpha_1}}{\partial t^{\alpha_1}}\Phi_x(\delta\hspace{0.5mm}t)|_{t=2(g+h)-\alpha_2-\alpha_3}.
\end{aligned}
\end{equation}
\noindent The proof can be found in \cite{jasour2021moment}. We can now utilize this result in order to propagate the disturbance through the agent's nonlinear dynamics in \eqref{eq:nonlin_dynamics}, and compute the moments of the uncertain states. The agent's dynamics in \eqref{eq:nonlin_dynamics} can equivalently be written in the following form:
\begin{equation} \label{eq:aug_dynamics}
	\hat{\boldsymbol{x}}_t = A_{t-1} \hat{\boldsymbol{x}}_{t-1} + B_{t-1},
\end{equation}
\noindent where $\hat{\boldsymbol{x}}_t = [x_t,y_t,z_t,\cos(\psi_t),\sin(\psi_t)]^\top$, and using the fact that:
\begin{equation}
\begin{aligned}
    \cos(\psi_t) &=  \cos(\psi_{t-1})\cos(\Delta_t(u^\psi_{t-1}+\omega^\psi_{t-1}))\\
    &\quad -\sin\psi_{t-1}\sin(\Delta_t(u^\psi_{t-1}+\omega^\psi_{t-1})) \\
	\sin(\psi_t) &=  \sin(\psi_{t-1})\cos(\Delta_t(u^\psi_{t-1}+\omega^\psi_{t-1}))\\
	&\quad +\cos(\psi_{t-1})\sin(\Delta_t(u^\psi_{t-1}+\omega^\psi_{t-1})).
\end{aligned}
\end{equation}
$A_t$ and $B_t$ are given by:
\begin{equation}
A_t =
\begin{bmatrix}
  	1 & 0 & 0 & \Delta_t(u^\nu_y\mathord{+}\omega^\nu_t) & 0 \\
    0 & 1 & 0 & 0 & \Delta_t(u^\nu_t\mathord{+}\omega^\nu_t) \\
    0 & 0 & 1 & 0 & 0 \\
    0 & 0 & 0 & \cos(\Delta_t(u^\psi_t\mathord{+}\omega^\psi_t)) & -\sin(\Delta_t(u^\psi_t\mathord{+}\omega^\psi_t)) \\
    0 & 0 & 0 & \sin(\Delta_t(u^\psi_t\mathord{+}\omega^\psi_t)) & \cos(\Delta_t(u^\psi_t\mathord{+}\omega^\psi_t)) \\
\end{bmatrix},
\end{equation}
and $B_t = [0,0, \Delta_t(u^z_t+\omega^z_t),0,0]^\top$. With this transformation, \eqref{eq:lemma_1} can be directly applied to \eqref{eq:aug_dynamics}, to compute the first moment $\boldsymbol{\mu}^1_t$ of the uncertain state $\hat{\boldsymbol{x}}_t$ in a recursive fashion as a linear combination of mixed-trigonometric-polynomial functions of the disturbance term $\boldsymbol{\omega}_t = [\omega^\nu_t, \omega^z_t, \omega^\psi_t]$, and the control inputs $\boldsymbol{u}_t = [u^\nu_t, u^z_t, u^\psi_t]$. To see this observe that $\boldsymbol{\mu}^1_t = \mathbb{E}[\hat{\boldsymbol{x}}_t]$ and so
 $\boldsymbol{\mu}^1_t = [\mathbb{E}[x_t], \mathbb{E}[y_t],\mathbb{E}[z_t], \mathbb{E}[\cos(\psi_t)], \mathbb{E}[\sin(\psi_t)]]^\top$. Subsequently, applying the expectation operator to \eqref{eq:aug_dynamics} we obtain the moment-state dynamics as:
 \begin{equation}\label{eq:first_moment}
\boldsymbol{\mu}^1_t = A^{mom_1}_{t-1} \boldsymbol{\mu}^1_{t-1} + B^{mom_1}_{t-1},
\end{equation}
where $A^{mom_1}_{t-1}$ and $B^{mom_1}_{t-1}$ are computed by the application of \eqref{eq:lemma_1} as:
\begin{equation}
A^{mom_1}_t =
\begin{bmatrix}
  	1 & 0 & 0 & \Delta_t u^\nu_t + \mu_{\Delta_t,\omega^\nu} & 0 \\
    0 & 1 & 0 & 0 & \Delta_t u^\nu_t + \mu_{\Delta_t,\omega^\nu} \\
    0 & 0 & 1 & 0 & 0 \\
    0 & 0 & 0 & \gamma_t & -\lambda_t \\
    0 & 0 & 0 & \lambda_t & \gamma_t \\
\end{bmatrix},
\end{equation}
where $\gamma_t = \cos(\Delta_t u^\psi_t)\mu_{\Delta_t,c_{\omega^\psi}}-\sin(\Delta_t u^\psi_t)\mu_{\Delta_t,s_{\omega^\psi}}$, $\lambda_t = \sin(\Delta_t u^\psi_t)\mu_{\Delta_t,c_{\omega^\psi}}+\cos(\Delta_t u^\psi_t)\mu_{\Delta_t,s_{\omega^\psi}}$, and finally $B^{mom_1}_t=[0,0,\Delta_t u^z_t + \mu_{\Delta_t,\omega^z},0,0]^\top$. The mixed moments of order  $\alpha$ i.e., $\boldsymbol{\mu}^\alpha_t = \mathbb{E}[(\hat{\boldsymbol{x}}_t)^\alpha] $ are obtained following the same procedure.

%The procedure described above allows us to compute the moments of the uncertain states of the agent for any random disturbance term which exhibits finite moments, by propagating a finite sequence of moments from the initial probability distributions through the nonlinear stochastic dynamical system. This results in the acquisition of the moments of the system's state probability distributions over a planning horizon. Subsequently, we can apply known inequalities from probability theory that offer bounds on the likelihood of a random variable falling within a specified range. Based on these bounds, we can design controllers that ensure the process remains within desired limits, maintaining a predetermined level of confidence.
\begin{thm}  
Let $x$ be a real valued unimodal random variable with finite expectation $\mathbb{E}[x]$, variance $\mathbb{E}[(x-\mathbb{E}[x])^2]$, and $r \ge 0$, the one-sided Vysochanskij-Petunin inequality is given by:
\begin{equation}\label{eq:VP_inequality}
P(x-\mathbb{E}[x]\geq r)\leq \frac{4}{9}\frac{\mathbb{E}[(x-\mathbb{E}[x])^2]}{r^2+\mathbb{E}[(x-\mathbb{E}[x])^2]},
\end{equation}
and holds when $r^2\geq\frac{5}{3}\mathbb{E}[(x-\mathbb{E}[x])^2]$. The proof can be found in \cite{VP}.
\end{thm}
First, we want to make sure that the agent reaches the region of interest $S_\kappa$, i.e., a sphere with center $\boldsymbol{\bar{s}}_\kappa = [s^\kappa_x, s^\kappa_y, s^\kappa_z]^\top$, and radius $s_r$, at the final state $\hat{\boldsymbol{x}}_{t+T^\prime|t}$ within a suitable selected planning horizon $T^\prime$. Let us denote the squared euclidean distance between the position of the UAV at the final state $\text{pos}(\hat{\boldsymbol{x}}_{t+T^\prime|t}) = [x_{T^\prime},y_{T^\prime},z_{T^\prime}]^\top$ of the horizon and the center of the region of interest $\boldsymbol{\bar{s}}_\kappa$ as $d^\kappa_{T^\prime} = ||\text{pos}(\hat{\boldsymbol{x}}_{t+T^\prime|t}) - \boldsymbol{\bar{s}}_\kappa||^2_2$. Because, the position of the UAV is uncertain, $d^\kappa_{T^\prime}$ becomes a random variable and therefore we are interested in computing the UAV control inputs such that: $P(d^\kappa_{T^\prime} \geq s_r^2) \leq \epsilon$ (i.e., the UAV is outside the region of interest with probability at most $\epsilon$) %$P(d^\kappa_T \leq S_r^2)$ is satisfied with some predetermined level of confidence. 
is satisfied with some predetermined level of confidence. To do that, we define $f^\kappa_{T^\prime} = s_r^2 - d^\kappa_{T^\prime}$ and we compute an upper bound on the probability $P(f^\kappa_{T^\prime} \leq 0)$ utilizing the Vysochanskij-Petunin inequality as:
\begin{equation} \label{eq:VP2}
P(f^\kappa_{T^\prime} \leq 0)\leq \frac{4}{9} \frac{\mathbb{E}[(f^\kappa_{T^\prime})^2]-\mathbb{E}[f^\kappa_{T^\prime}]^2}{\mathbb{E}[f^\kappa_{T^\prime}]^2},
\end{equation}
\noindent which is valid when $\mathbb{E}[f^\kappa_{T^\prime}]\geq 0$ and $\mathbb{E}[f^\kappa_{T^\prime}]^2 \geq \frac{5}{8}\mathbb{E}[(f^\kappa_{T^\prime})^2]$. This result has been obtained by applying \eqref{eq:VP_inequality} to the unimodal random variable $x = -f^\kappa_{T^\prime}$, and setting $ r= \mathbb{E}[f^\kappa_{T^\prime}]$. The inequality in \eqref{eq:VP2} establishes the upper bound on $P(f^\kappa_{T^\prime} \leq 0)$, as a function of $\mathbb{E}[f^\kappa_{T^\prime}]$ and of $\mathbb{E}[(f^\kappa_{T^\prime})^2]$. By expanding $f^\kappa_{T^\prime}$ and $(f^\kappa_{T^\prime})^2$, and applying the expectation operator on the result, the upper bound can be expressed in terms of the first four moments of the system states. For instance, $\mathbb{E}[f^\kappa_{T^\prime}]$ is computed as $\mathbb{E}[ (x_{T^\prime}-s^\kappa_x)^2 + (y_{T^\prime}-s^\kappa_y)^2 + (z_{T^\prime}-s^\kappa_z)^2 - s_r^2]$ which is equal to:
\begin{align} \label{eq:avg}
&\mathbb{E}[(x_{T^\prime})^2]-2\mathbb{E}[x_{T^\prime}]s^\kappa_x+(s^\kappa_x)^2+ \notag \\
&\mathbb{E}[(y_{T^\prime})^2]-2\mathbb{E}[y_{T^\prime}]s^\kappa_y+(s^\kappa_y)^2+ \notag \\
&\mathbb{E}[(z_{T^\prime})^2]-2\mathbb{E}[z_{T^\prime}]s^\kappa_z+(s^\kappa_z)^2-s_r^2.
\end{align}
\noindent The first moments of the UAV state shown in the example above have been computed recursively through the moment-state dynamics as shown in \eqref{eq:first_moment}, which are exclusively function of the control inputs. Required higher order moments (in our case of order up to $\alpha=4$ which are required by $\mathbb{E}[(f^\kappa_{T^\prime})^2]$) are computed in a similar fashion.

\begin{figure*}
	\centering
	\includegraphics[width=\textwidth]{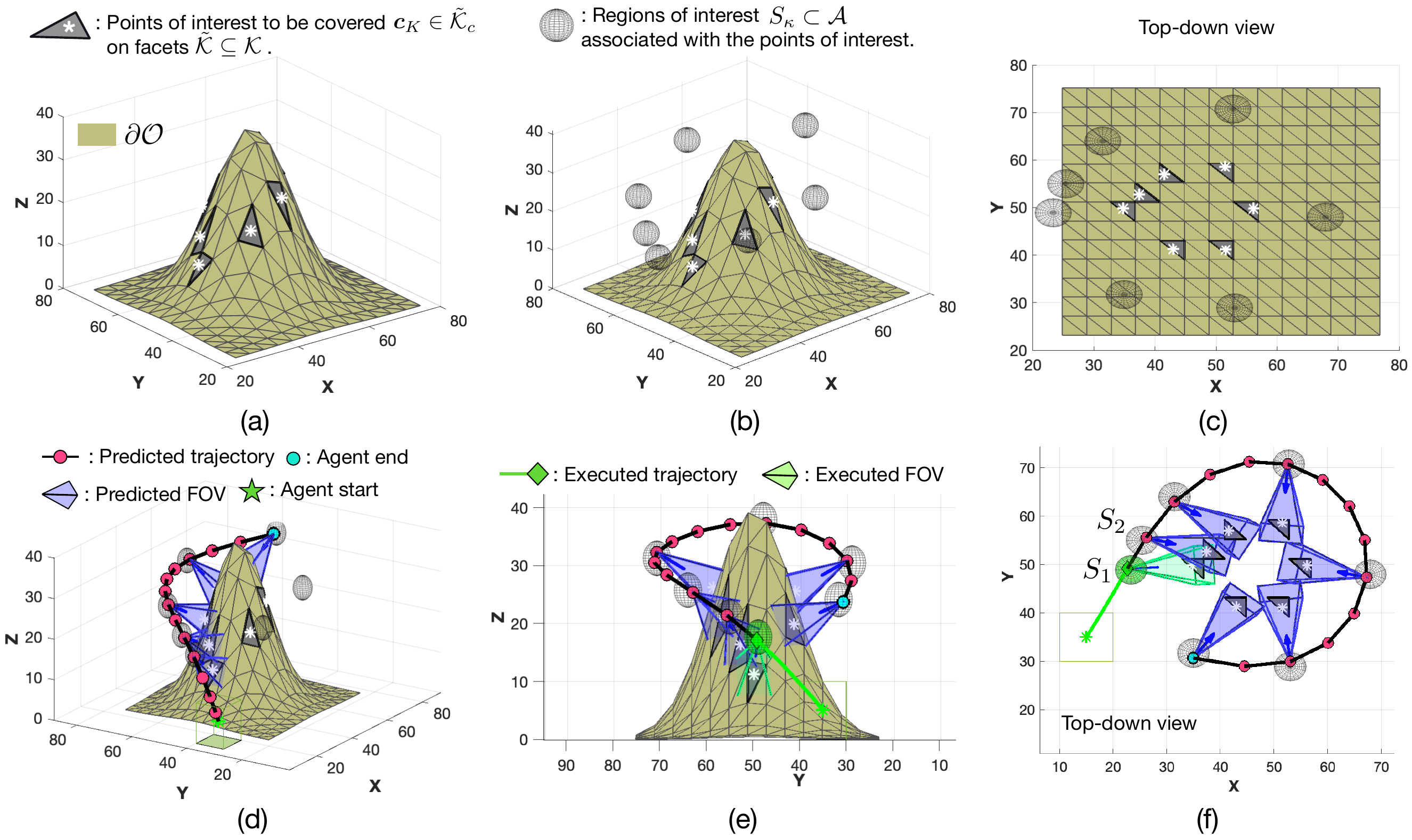}
	\caption{The figure shows an illustrative example of the stage-1 controller.}	
	\label{fig:fig1}
	\vspace{-0mm}
\end{figure*}

Finally, the stage-2 controller solves the optimization problem shown in \eqref{eq:P3} to robustly guide the agent inside $S_\kappa$ with probability at least $1 - \epsilon$.

\begin{algorithm}[h!]
\begin{subequations} \label{eq:P3}
\begin{align} 
%&\hspace*{-3mm}\texttt{Stage-1 MPC Controller} & \notag\\
& \hspace*{-2.5mm} ~~\underset{\{\boldsymbol{u} \in \mathcal{U}\}} {\min} ~\mathcal{J}^{0} = \sum_{t=1}^T ||\boldsymbol{u}_{t-1}||_2^2 &  \hspace*{-5mm} \label{eq:P3_0} \\
&\hspace*{-3mm}\textbf{subject to: $\tau \in \{0,..,{T^\prime}-1\}, n \in \{1,..,4\}$} ~  &\nonumber\\
&\hspace*{-3mm} \boldsymbol{\mu}^n_{t+\tau+1|t} = A^{mom_n}_{t+\tau|t} \boldsymbol{\mu}^n_{t+\tau|t} + B^{mom_n}_{t+\tau|t}, & \hspace*{-5mm} \forall n, \tau \label{eq:P3_1}\\
&\hspace*{-3mm} \boldsymbol{\mu}^n_{t|t} = \boldsymbol{\mu}^n_{t-1|t}, & \hspace*{-5mm} \forall n, \tau \label{eq:P3_2}\\
&\hspace*{-3mm} \frac{4}{9} \frac{\mathbb{E}[(f^\kappa_{t+{T^\prime}|t})^2]-\mathbb{E}[f^\kappa_{t+{T^\prime}|t}]^2}{\mathbb{E}[f^\kappa_{t+{T^\prime}|t}]^2} \leq \epsilon, & \hspace*{-5mm}  \label{eq:P3_3}
%&\hspace*{-3mm} \frac{4}{9} \frac{\mathbb{E}[(\sigma^\kappa_{t+T|t})^2]-\mathbb{E}[\sigma^\kappa_{t+T|t}]^2}{\mathbb{E}[\sigma^\kappa_{t+T|t}]^2} \leq \epsilon/2, & \hspace*{-5mm}  \label{eq:P3_4}\\
%&\hspace*{-3mm} \frac{4}{9} \frac{\mathbb{E}[(\xi^\kappa_{t+T|t})^2]-\mathbb{E}[\xi^\kappa_{t+T|t}]^2}{\mathbb{E}[\xi^\kappa_{t+T|t}]^2} \leq \epsilon/2, & \hspace*{-5mm}  \label{eq:P3_5}
%&\hspace*{-3mm} n = \{1,..,4\} & \hspace*{-7mm}
\vspace{-3mm}
\end{align}
\end{subequations}
\end{algorithm}

\section{Evaluation} \label{sec:eval}

%(a) Shows the admissible FOV configurations that can be executed by the gimbal, (b) shows the object of interest, and the points of interest to be covered, (c)-(d) shows the reference coverage plan generated by the stage-1 controller

%\begin{figure}
%	\centering
%	\includegraphics[width=\columnwidth]{figs/fig2b.pdf}
%	\caption{The figure illustrates the orientation control executed by the stage-2 controller, facilitating coverage of the point of interest associated with the region of interest $S_2$. It robustly aligns the agent's orientation ($\psi_{t+T|t}$) with $\psi\kappa$, selected close to the direction of the reference plan (indicated by the red arrow, representing the tangent) while ensuring the existence of a FOV configuration capable of covering the point of interest.}	
%	\label{fig:fig2}
%	\vspace{-0mm}
%\end{figure}

\subsection{Simulation Setup}

To demonstrate the proposed approach we have used the following simulation setup: the agent dynamics are given by \eqref{eq:nonlin_dynamics} with horizontal and vertical velocities set as $u_t^\nu \in [0,10] ~\text{m}/\text{s}$, and $u_t^z \in [-10,10] ~\text{m}/\text{s}$ respectively. The yaw rate is given by $u_t^\psi \in [-\pi,\pi] ~\text{rad}/\text{s}$, and the sampling interval is $\Delta_t=0.1$s. The disturbance $\boldsymbol{\omega}_t = [\omega^\nu_t, \omega^z_t, \omega^\psi_t]^\top$ is defined as follows: $\omega^\nu_t$ follows a Beta distribution with parameters $\alpha = 1$, and $\beta = 3$, $\omega^z_t$ follows a zero-mean Gaussian distribution $\mathcal{N}(0,0.3) ~\text{m}/\text{s}$, and finally, $\omega^\psi_t$ follows a Uniform distribution in the range $[-0.1,0.1] ~\text{rad}/\text{s}$. The agent camera FOV model parameters $(h_f, w_f, \ell_f)$ are set to $(16, 8, 8)$m, and the gimbal rotation angles $\phi^y$ and $\phi^z$ take their values from the finite sets $\Phi^y=\{\frac{\pi}{8}, \frac{\pi}{4}, \frac{3\pi}{8}, 0, -\frac{\pi}{8}, -\frac{\pi}{4}, -\frac{3\pi}{8}\}~\text{rad}$, and  $\Phi^z=\{\frac{\pi}{4},0,-\frac{\pi}{4}\}~\text{rad}$, respectively leading to $|\mathcal{M}|=21$ possible camera FOV configurations. Finally, we should mention that the stage-1 controller in \eqref{eq:P2} was implemented as a mixed integer quadratic program (MIQP), and solved using the Gurobi solver, whereas the stage-2 controller in \eqref{eq:P3} was solved using off-the-shelf non-linear optimization tools.

\subsection{Results}

An illustrative example of the proposed approach is shown in Fig. \ref{fig:fig1}. Specifically, Fig. \ref{fig:fig1}(a) shows the object of interest $O$ to be covered with its surface area $\partial O$ given by the function $q(x,y) = 40\exp\left(-\left(\frac{(x-45)^2}{160}+\frac{(y-45)^2}{160}\right)\right)$, triangulated into a 3D mesh composed of $338$ triangular facets $K \in \mathcal{K}$. A subset $\tilde{\mathcal{K}}$ of those facets have been randomly sampled, and their centroids $\boldsymbol{c}_K \in \tilde{\mathcal{K}}_c$ have been marked for coverage. In total,  $|\tilde{\mathcal{K}}_c| = 7$ points of interest need to be covered as shown in Fig. \ref{fig:fig1}(a). For each point of interest $\boldsymbol{c}_K \in \tilde{\mathcal{K}}_c$, we have associated a region of interest $S_\kappa$ ($\ell=12 \text{m}, s_r=3\text{m}$), where $\kappa$ is the index to the associated point $\boldsymbol{c}_K$ of facet $K$, also shown in Fig. \ref{fig:fig1}(b)-(c). 
\begin{figure}
	\centering
	\includegraphics[scale=0.55]{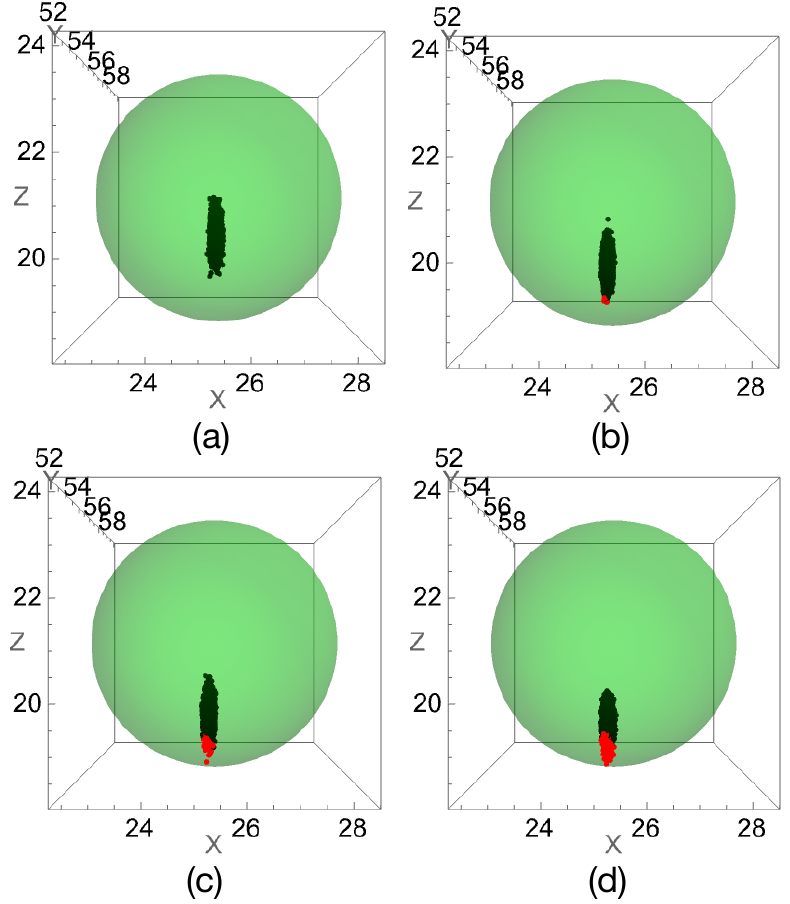}
	\caption{The figure illustrates a Monte-Carlo simulation of the agent's state (i.e., position) at the end of the horizon obtained by the stage-2 controller for different confidence levels $\epsilon$. (a) $\epsilon=0.005$, (b) $\epsilon=0.0250$, (c) $\epsilon=0.05$, and (d) $\epsilon=0.1$.}	
	\label{fig:fig3}
	\vspace{-0mm}
\end{figure}
Then, Fig. \ref{fig:fig1}(d)-(f) shows the reference coverage plan generated by the model-predictive stage-1 controller discussed in Sec. \ref{ssec:stage1}, by optimizing \eqref{eq:P2_0} with $w=10^{-3}$. In particular, Fig. \ref{fig:fig1}(d) illustrates the agent's trajectory inside a rolling planning horizon of length $T=14$ time-steps, generated according to the high-level dynamics shown in \eqref{eq:P2_1} with $\Delta^1_t=1 \text{s}$, and $|\boldsymbol{u}^1_t| \leq 10 \text{m}/\text{s}$. In this experiment, we select FOV configurations from the set $\mathcal{M}$. As demonstrated in Fig. \ref{fig:fig1}(d), the trajectory $\boldsymbol{x}^1_{1+\tau+1|1}$, for $\tau \in \{1,..,14\}$, is guided through the regions of interest, selecting FOV configurations that result in the coverage of the points of interest by optimizing \eqref{eq:P2_0}. Similarly, Fig. \ref{fig:fig1}(e)-(f) displays the generated reference plan $\boldsymbol{x}^1_{6+\tau+1|6}$ for $\tau \in \{1,..,14\}$, with the executed coverage plan marked green.

As we have already discussed, at every time-step the stage-1 controller generates the coverage reference plan, and sends the region of interest to be visited next to the stage-2 controller. 

Subsequently, the stage-2 controller aims to: robustly guide the agent inside the region of interest. This is demonstrated in next, Fig. \ref{fig:fig3} with a Monte-Carlo simulation of the output of stage-2 controller during the transition of the agent from $S_1$ to $S_2$, (as shown in Fig. \ref{fig:fig1}(f)) over a planning horizon of $T^\prime=14$ time-steps, for different confidence levels $\epsilon$. The figure shows the agent's position at the end of the planning horizon. Specifically, Fig. \ref{fig:fig3}(a) shows 10000 samples of the agent's final position for $\epsilon=0.005$. As shown, all samples of the agent's position reside inside the region of interest indicated by the green sphere (i.e., $S_2$). Similarly, Fig. \ref{fig:fig3}(b), Fig. \ref{fig:fig3}(c), and Fig. \ref{fig:fig3}(d) shows the same result for $\epsilon=0.0250$, $\epsilon=0.05$, and $\epsilon=0.1$ respectively. As shown in the figure, the red points indicate samples violating the constraints but always within the selected confidence level. In Fig. \ref{fig:fig3}(b), 4 samples (out of 10000) reside outside the region of interest, well below the desired $\epsilon$ which is un upper bound of the violation probability. The same applies for Fig. \ref{fig:fig3}(c) and Fig. \ref{fig:fig3}(d) where 56 and 377 samples violate the constraints respectively.

%\begin{figure*}
%	\centering
%	\includegraphics[width=\textwidth]{figs/fig3.pdf}
%	\caption{TODO}	
%	\label{fig:fig3}
%	\vspace{-0mm}
%\end{figure*}

%\vspace{-0.2mm}
\section{Conclusion} \label{sec:conclusion}
We introduce a two-stage hierarchical fault-tolerant coverage controller for UAVs to cover 3D points of interest. The first stage formulates a hybrid optimal control problem, optimizing UAV mobility and camera control inputs for maximal coverage. The second stage employs mixed-trigonometric-polynomial moment propagation, to generate fault-tolerant coverage trajectories by accommodating non-Gaussian disturbances on the control inputs propagated through the system's nonlinear dynamics. Future work will explore fault-tolerant camera control and fault-tolerant multi-agent coverage control.

%==============================================================================

\bibliography{ifacconf}             
                                                     
\end{document}